\documentclass[aps,prb,groupedaddress]{revtex4}
\usepackage[dvips]{graphicx}

\begin{document}

\title{Spin-torque switching: Fokker-Planck rate calculation}
\author{D. M. Apalkov}
\email{apalk001@ua.edu}
\author{P. B. Visscher}
\email{visscher@ua.edu}
\homepage{http://bama.ua.edu/~visscher/mumag}
\affiliation{MINT Center and Department of Physics and Astronomy\\
University of Alabama, Tuscaloosa AL 35487-0324}
\date{\today}

\begin{abstract}
We describe a new approach to understanding and calculating
magnetization switching rates and noise in the recently observed
phenomenon of "spin-torque switching".  In this phenomenon, which
has possible applications to information storage, a large current
passing from a pinned ferromagnetic (FM) layer to a free FM layer
switches the free layer. Our main result is that the spin-torque
effect increases the Arrhenius factor $\exp(-E/kT)$ in the
switching rate, not by lowering the barrier $E$, but by raising
the effective spin temperature $T$. To calculate this effect
quantitatively, we extend Kramers' 1940 treatment of reaction
rates, deriving and solving a Fokker-Planck equation for the
energy distribution including a current-induced spin torque of the
Slonczewski type. This method can be used to calculate slow
switching rates without long-time simulations; in this Letter we
calculate rates for telegraph noise that are in good qualitative
agreement with recent experiments.  The method also allows the
calculation of current-induced magnetic noise in CPP (current
perpendicular to plane) spin valve read heads.
\end{abstract}

\maketitle



\section{Introduction}

Recently it has been demonstrated that the magnetization of a thin
ferromagnetic film can be switched by passing a current between it
and a pinned layer\cite{albert}.  This "spin-torque switching"
phenomenon is of interest for possible information storage
applications. Except at very high currents, the switching appears
to be thermal in nature.
Previous theoretical treatments of thermal spin-torque switching\cite%
{myers,zhang,zhangXXX} have been based on the idea that the spin
torque increases the rate by lowering the effective potential
energy barrier, and have encountered a fundamental problem: the
common Slonczewski\cite{slon96,slon99} model for the spin torque
is not conservative, so it cannot be described by a potential
energy. %
The effects of the Slonczewski torque on the Landau-Lifshitz (LL)
equation for the magnetization dynamics are similar to those of
the LL damping, so in our Fokker-Planck approach it makes a
contribution to the effective damping.  When this contribution is
negative, the effective temperature is raised.  The notion of an
elevated effective temperature during spin-torque switching has
been discussed previously\cite{urazhdin,wegrowe,koch}; the present
Fokker-Planck formulation allows the precise definition and
calculation of the effective temperature, which we will refer to
as the Maxwell-Boltzmann temperature (Eq. \ref{drho}) and
clarifies the relation between it and the (lower) LL noise
temperature.

The Fokker-Planck equation gives the time evolution of a phase
space probability density$.$\ It was first applied to chemical
rate problems in 1940 by Kramers\cite{kramers}, who observed that
except for very large or very small damping constants, the escape
rate is well described by an earlier ''transition state theory''
(TST)\cite{eyring}, in which the rate of barrier-crossing in a
non-equilibrium system is assumed to be the same as that in an
equilibrium system. \ Although corrections to TST have been
extensively studied\cite{chandler,visscher}, TST has been found to
be the most useful starting point for rate calculations. \ In this
Letter we will use a TST-like approximation, differing from the
usual TST in that the system is not in a true thermal equilibrium,
but a non-equilibrium steady state. \ We will write the magnetic
Fokker-Planck equation of Brown\cite{brown}, generalized to
include the Slonczewski torque, but following
Kramers\cite{kramers} convert it to describe diffusion in energy
rather than magnetization; to the best of our knowledge this has
not been done previously except for systems with azimuthal
symmetry\cite{brown,coffey}. \

The LL equation\cite{general} for the evolution of a uniform magnetization $%
\mathbf{M}(t)$ has a deterministic and a random part:
\begin{equation}
\dot{\mathbf{M}}{\equiv }\frac{d{{\mathbf{M}}}}{dt}=\dot{\mathbf{M}}_{\text{%
det}}+\dot{\mathbf{M}}_{\text{rand}}  \label{LL}
\end{equation}%
The deterministic part is divided into a conservative precession
term and the dissipative LL damping, and we will include also the
Slonczewski
current-induced torque:%
\begin{equation}
\dot{\mathbf{M}}_{\text{det}}=\dot{\mathbf{M}}_{\text{cons}}+\dot{\mathbf{M}}%
_{\text{LL}}+\dot{\mathbf{M}}_{\text{Slon}}  \label{Mdet}
\end{equation}%
We will first specify the precession torque:%
\begin{equation}
\dot{\mathbf{M}}_{\text{cons}}=-\gamma \mathbf{M}\times \mathbf{H}_{\text{%
cons}}  \label{Mcons}
\end{equation}%
where $\gamma $ is the gyromagnetic ratio. \ We refer to the field $\mathbf{H%
}_{\text{cons}}$ about which $\mathbf{M}$ precesses as
''conservative'' because it can be written as the gradient with
respect to $\mathbf{M}$ of an energy density,
\begin{equation}
\mu _{0}\mathbf{H}_{\text{cons}}=-\nabla E(\mathbf{M)}.
\label{gradd}
\end{equation}%
(This is a 2D gradient on the $\mathbf{M}$-sphere; see Eq.
\ref{A.Hcons} of ref. \cite{epaps}.)

\begin{figure}[tbh]
\begin{center}
\includegraphics[height=1.3 in]{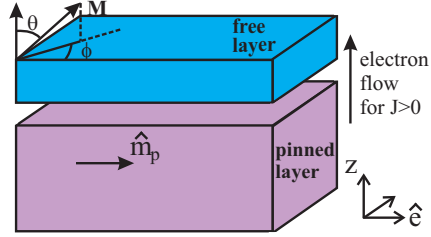}
\end{center}
\par
\caption{Geometry of thin-film element, for the case where the
magnetization
$\hat{\mathbf{m}}_{\text{p}}$ of the "pinned" layer is along the easy axis $%
\hat{\mathbf{e}}$ of the free layer. } \label{geom}
\end{figure}

Our derivation of the FP equation is valid for a system with
arbitrary anisotropy, but for specificity we will consider the
case of a thin-film element (Fig. \ref{geom})
for which the energy density is given (in SI units)
by\cite{general}
\begin{equation}
E(\mathbf{M})/\mu _{0}=-\frac{1}{2}H_{K}M_{s}\mathbf{{(\hat{{m}}\cdot \hat{e}%
)}}^{2}+\frac{1}{2}M_{s}^{2}\mathbf{{(\hat{{m}}\cdot \mathbf{\hat{z}})}}^{2}-%
\mathbf{{{H}_{\text{ext}}\cdot \mathbf{M}}}  \label{E}
\end{equation}%
Here $\mathbf{{{H}_{\text{ext}}}}$ is an external field, $H_{K}$
is the uniaxial anisotropy
field, $\mathbf{\hat{m}\equiv M}/M_{s}$, $\mathbf{\hat{e}}$, and $\mathbf{%
\hat{z}}$ are unit vectors along the magnetization, easy axis, and
z axis (perpendicular to the film) respectively, and $M_{s}$ is
the saturation magnetization. \

The nonconservative LL damping torque (Eq. \ref{Mdet})
is\cite{general}
\begin{equation}
\dot{\mathbf{M}}_{\text{LL}}\mathbf{=-}\gamma \alpha M_{s}\hat{\mathbf{m}}%
\times (\hat{\mathbf{m}}\times \mathbf{H}_{\text{cons}})=\gamma \alpha M_{s}%
\mathbf{H}_{\text{cons}}  \label{MLL}
\end{equation}%
where $\alpha $ is the dimensionless LL damping constant. \
The Slonczewski spin-torque\cite%
{slon99,zhang} is
\begin{equation}
\dot{\mathbf{M}}_{\text{Slon}}=-\gamma JM_{s}\hat{\mathbf{m}}\times (\hat{%
\mathbf{m}}\times \hat{\mathbf{m}}_{\text{p}})  \label{MSlon}
\end{equation}%
where $J$ is an empirical constant with units of magnetic field,
proportional to the current density, and
$\mathbf{{\hat{{m}}}_{\text{p}}}$ is the magnetization direction
in the thicker (often pinned) layer from (or to) which the current
flows.

\begin{figure}[tbh]
\begin{center}
\includegraphics[height=2.0 in]{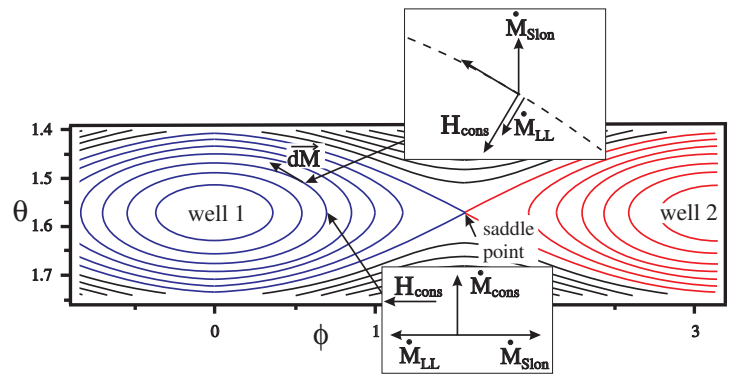}
\end{center}
\par
\vspace{-0.2 in} 
\caption{Energy contours (Stoner-Wohlfarth orbits) for a thin
film, plotted in terms of the coordinates $\protect\theta $ and
$\protect\phi $ defined in
Fig. \ref{geom}, for the case $H_{K}/M_{s}{=0.028}$, $\mathbf{{{H}_{\text{%
ext}}}}=0.$\ The regions $i=1,2,3$ are blue, red, and black
respectively.  The vertical scale is exaggerated for clarity. \
Lower inset: contributions to the rate of change of magnetization
for a magnetization in the film plane. \ The insets show the
tangent plane: magnetization points out of the paper. \ Upper
inset: the same for an arbitrarily chosen direction of
$\mathbf{M}$.} \label{cont}
\end{figure}

The effect of the random torque $\dot{\mathbf{M}}_{\text{rand}}$
is to produce a diffusive random walk on the surface of the
$\mathbf{M}$-sphere. \ We will relate this to a diffusivity $D$
(Eq. \ref{fluctdis}) by
giving the mean square value of the increment $\Delta \mathbf{M}_{\text{rand}%
}=\dot{\mathbf{M}}_{\text{rand}}\Delta t$:%
\begin{equation}
\left\langle \Delta \mathbf{M}_{\text{rand}}^{2}\right\rangle%
=4D\Delta t .
\end{equation}%
The directions of these torques are shown in insets to Fig.
\ref{cont}, from which the basic mechanism of spin-torque
switching can be seen: the Slonczewski torque pulls the
magnetization out of well $1$ and allows it to jump to well $2$.

The Fokker-Planck equation describes the evolution of a probability density %
$\rho (\mathbf{M},t)$ on the $\mathbf{M}$-sphere.  It can be
written in the form of a continuity equation\cite{brown} for %
$\rho (\mathbf{M},t)$: %
\begin{equation}
\frac{\partial \rho (\mathbf{M},t)}{\partial t}=-\nabla \cdot \mathbf{j}(%
\mathbf{M},t)  \label{contin}
\end{equation}%
where the probability current $\mathbf{j}$ along the sphere has a
convective and a diffusive part:
\begin{equation}
\mathbf{j}(\mathbf{M},t)\equiv \rho (\mathbf{M},t)\dot{\mathbf{M}}_{\text{det%
}}(\mathbf{M})-D\nabla \rho (\mathbf{M},t)  \label{j}
\end{equation}%
(note that both the divergence and the gradient are
two-dimensional here). \ Inserting Eq. \ref{j} into Eq.
\ref{contin} gives the FP equation (Eq. \ref{A.FP} of ref.
\cite{epaps}) first derived (without the spin torque term) in 1963
by Brown\cite{brown}.

Frequently, the probability density depends mostly on energy,
being constant along an orbit and depending weakly on phase around
the orbit. \ This is exactly true in a thermal equilibrium system
(even with damping), and we show below that it is true in a steady
state system with a Slonczewski torque, modeling the telegraph
noise system. \ It has often been assumed to be approximately true
away from the barrier, to compute non-equilibrium switching
rates\cite{kramers,brown,coffey}. \ The energy dependence may be
different in different regions of the sphere (for example,
different energy wells), so we will define a density $\rho
_{i}^{\prime
}(E,t)$, where the region (well) index $i=1$ for the $\phi =0$ well (for $%
E\leq E_{\text{sad}}$), $i=2$ for the $\phi =\pi $ well (Fig. \ref{cont}), and $%
i=3$ for $E\geq E_{\text{sad}}$. \ [The three will be equal at the
saddle point, where all three regions touch.] \ This density $\rho
^{\prime }$ is related to $\rho_i $ by

\begin{equation}
\rho (\mathbf{M},t)=\rho _{i}^{\prime }(E(\mathbf{M)},t)
\label{rhoE}
\end{equation}

Kramers derived a Fokker-Planck equation in energy for a particle
in a well, but we are not aware of any previous derivation for the
magnetic case so we will derive it here. Though Kramers used it
only in the low-damping limit, it is an exact description of the
steady state of the system even for high damping.

The FP equation in energy takes the form of a continuity equation
\begin{equation}
\frac{\gamma M_{s}P_{i}(E)}{\mu _{0}}\frac{\partial \rho _{i}^{\prime }(E,t)%
}{\partial t}=-\frac{\partial }{\partial E}j_{i}^{E}(E,t)
\label{Econtin}
\end{equation}%
where the current $j_{i}^{E}(E,t)$ is the number of systems per
unit time crossing a constant-energy contour. \ There is a factor
on the left hand side involving the orbital period $P_{i}(E)$
because $\rho ^{\prime }$ is not the probability per unit energy
but per unit area on the $M$-sphere (see Eq. \ref{A.DOS} of ref.
\cite{epaps}). \ The current in energy can be obtained from the
current on the $M$-sphere (Eq. \ref{j}; see Eqs.
\ref{A.jE}-\ref{A.jdiff} of ref. \cite{epaps} for details):
\begin{equation}
j_{i}^{E}(E,t)=-\gamma \alpha M_{s}\rho_i ^{\prime }(E,t)I_{i}(E)+\gamma JM_{s}%
\rho_i ^{\prime }(E,t)\hat{\mathbf{m}}_{\text{p}}\cdot \mathbf{I}_{i}^{M}-D%
\frac{\partial \rho_i ^{\prime }(E,t)}{\partial E}I_{i}(E)
\label{jE}
\end{equation}%
in terms of a damping term involving an energy integral over an
orbit in the $i^{\text{th}}$ well
\begin{equation}
I_{i}^{E}(E)\equiv \oint H_{\text{cons}}dM , \label{IE}
\end{equation}%
a Slonczewski torque term involving a magnetization integral%
\begin{equation}
\mathbf{I}_{i}^{M}(E,t)=\oint d\mathbf{M}\times \mathbf{M}
\label{IM}
\end{equation}%
and a diffusion term.

\begin{figure}[tbh]
\begin{center}
\includegraphics[height=2.0 in]{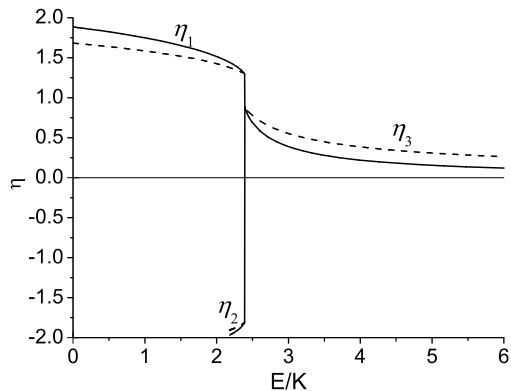}
\end{center}
\par
\vspace{-0.2 in} 
\caption{The current-damping coefficients $\protect\eta _{1}$ and $\protect%
\eta _{2}$ in the two potential wells, and $\protect\eta _{3}$ for
the region above the saddle-point energy. The values we actually
use (near the bottoms of the wells) depend only weakly on the
parameters $H_{\text{ext}}=-120$ Oe and $H_K=220$ Oe, which were
estimated from the fit to experimental data (Eq. \ref{A.Hk} of
ref. \cite{epaps}).
The averages $\overline{\protect\eta }%
_{i}$ (Eq. \ref{etabar}) are also shown, as dashed
lines.}\label{eta}
\end{figure}

For the telegraph-noise problem we require the steady-state form
obtained by setting $j_{i}^{E}=0$:
\begin{equation}
\frac{\partial \ln \rho ^{\prime }(E)}{\partial E}=\frac{\gamma
}{D}\left[ -\alpha M_{s}+\eta (E)J\right] \equiv -V\beta (E)
\label{drho}
\end{equation}%
where the right hand side defines an effective inverse
"Maxwell-Boltzmann" temperature $\beta (E) $, and $V$ is the
volume of the switching element. We have also defined a
dimensionless spin-torque-damping ratio $\eta (E)$ (Fig.
\ref{eta}) as the ratio of the work of the Slonczewski torque (Eq.
\ref{MSlon}) to that of the LL damping (Eq. \ref{MLL})
\begin{equation}
\eta (E)=\frac{\hat{\mathbf{m}}_{\text{p}}\cdot \mathbf{I}_{i}^{M}(E)}{%
I_{i}^{E}(E)}. \label{etadef}
\end{equation}%
Eq. \ref{drho} shows clearly that the Slonczewski torque acts like
a correction to the LL damping $\alpha$. \ Because $\eta $ has
opposite signs in the two wells, the damping contribution is
negative in one well and positive in the other. \ A similar result
has been suggested previously\cite{koch} for the special case in
which $\hat{\mathbf{m}}_{\text{p}}$ is parallel to
$\mathbf{H_{\text{ext}}}$.

If $J=0$, we get the expected Boltzmann
distribution with $\beta =1/k_{B}T$ only if%
\begin{equation}
D=\gamma M_{s}\alpha k_{B}T/V;  \label{fluctdis}
\end{equation}%
this is the fluctuation-dissipation theorem.

If we integrate Eq. \ref{drho} downward from the saddle point into
well $i=1$
or 2, we get%
\begin{equation}
\rho _{i}^{\prime }(E)=\rho ^{\prime }(E_{\text{sad}})\exp \left[ \frac{V}{%
k_{B}T}[1-\overline{\eta _{i}}(E)J/\alpha
M_{s}][E_{\text{sad}}-E]\right] \label{rho}
\end{equation}%
where the average $\eta$ is
\begin{equation}
\overline{\eta }_{i}(E)\equiv \frac{1}{[E_{\text{sad}}-E]}%
\int\limits_{E}^{E_{sad}}\eta _{i}(E^{\prime })dE^{\prime }
\label{etabar}
\end{equation}%
In region $i=3$ we must integrate upwards from $E_{\text{sad}}$
(see Eq. \ref{A.rho3} of ref. \cite{epaps}). \ The ratio $\eta $
and its average $\overline{\eta }$ are weakly dependent on $E$
(Fig. \ref{eta}), so the distribution is nearly a Boltzmann
distribution with an effective temperature $T/[1-\overline{\eta
}(E)J/\alpha M_s]$.

We now compute the switching rates using transition state theory
(TST). The TST rate is the steady-state probability per unit time
of crossing a vertical line ($\phi = \pi /2$) through the saddle
point in Fig. \ref{cont}.  This gives\cite{epaps}
\begin{equation}
j_{\text{TST}}=\frac{\gamma M_{s}k_{B}T}{\mu _{0}V[1-\overline{\eta _{3}}(E_{%
\text{sad}})J/\alpha M_{s}]}\rho ^{\prime }(E_{\text{sad}})
\label{jTST}
\end{equation}%

\begin{figure}[tbh]
\begin{center}
\includegraphics[height=3.0 in]{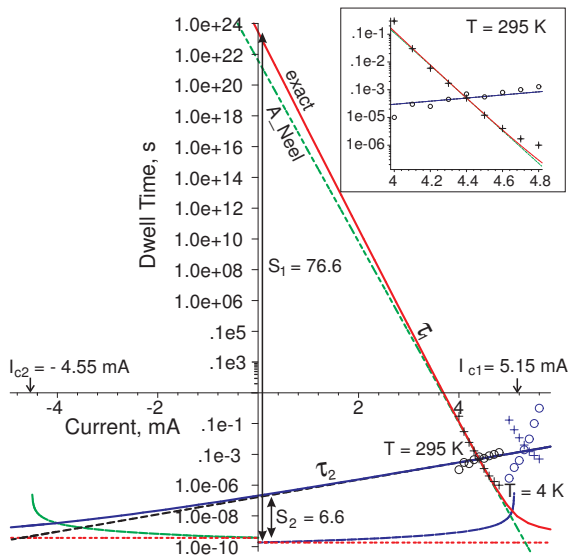}
\end{center}
\par
\vspace{-0.2 in} 
\caption{Dwell times as functions of current, illustrating the
fitting of the room-temperature data of %
Urazhdin \textit{et al}\protect\cite{urazhdin}. \ %
Inset (upper right) shows the experimental points ($+$ for
$\tau_1$, o for $\tau_2$) more clearly. The exact results are
solid lines, the Arrhenius-Neel approximation is dotted. \ Dotted
horizontal lines at the bottom are the orbit periods ($P_{1\text{
}}$ on the right, $P_{2\text{ }}$ on the left). \ Solid curves
that approach them at $I=0$ are the prefactor $P_{i\text{ }}
(1-I/I_{c3}) /(1-I/I_{ci})$, which intersects the exact
$\protect\tau _{i}$ at $I_{ci}$. We use $I_{c3}=I_{c1}
\overline{\eta_1} / \overline{\eta_3}$. } \label{dwell}
\end{figure}

From the $\rho _{i}^{\prime }$s (Eq. \ref{rho}) it is
straightforward to obtain the total probability $p_{i}$ of being
in each well (Eq. \ref{A.p} of ref. \cite{epaps}). \ With the
absolute-rate-theory current $j_{\text{TST}}$ (Eq. \ref{jTST})
these determine the dwell times $\tau _{1}$ and $\tau _{2\text{ }}$.%
We will write these in terms of a stability factor
$S_{i}\equiv VE^b_{i}/k_{B}T$ ($E^b_i$ is the barrier height
$E_{\text{sad}}-E_{i}$, where $E_{i}$ is the bottom of well $i=1$
or $2$) and a critical current at which the exponent in Eq.
\ref{rho} vanishes, %
$J_{ci}\equiv \alpha M_s / \overline{\eta _{i}}(E_{i})$. \ %
Since we do not know the exact proportionality factor between the
parameter $J$ and the actual physical current $I$ we can write
$J/J_{ci}$ as $I/I_{ci}$, where the critical currents $I_{ci}$
should be related by
\begin{equation}
I_{c1}\overline{\eta _{1}}(E_{1})=I_{c2}\overline{\eta_{2}}(E_{2})
 = I_{c3} \overline{\eta_{3}}(E_{\text{sad}}) \label{Icratio}
\end{equation}
\ Then the dwell times are given by\cite{epaps}
\begin{equation}
\tau _{i}=\frac{p_{i}}{j_{\text{TST}}}=P_{i}(E_{i})\frac{1-I/I_{c3}}{%
1-I/I_{ci}}\left[ e^{S_{i}(1-I/I_{ci})}-1\right]   \label{dwelli}
\end{equation}%
\ We define an ''Arrhenius-Neel'' approximation by neglecting $I$
in the prefactor and the $-1$:%
\begin{equation}
\tau _{i}^{A-N}=P_{i}(E_{i})e^{S_{i}(1-I/I_{ci})}
\end{equation}%
so that the dwell time is just a straight line on a logarithmic plot of $\tau$ (Figure %
\ref{dwell}). \ We adjust the two parameters $S_{1}$ and $I_{c1}$\
to match the slope and value of the measured\cite{urazhdin} dwell
time at the current $I=4.4$ ma at which $\tau _{1}$ and $\tau
_{2}$ cross. \ In the Arrhenius-Neel approximation, these
constants have simple graphical interpretations: $I_{c1}$ is the
current at which $\tau _{1}$ intersects the horizontal line at the
prefactor $P_{i}$ (the orbit period), and $S_{1}$ is the
(logarithmic) height of the dwell time above this prefactor at
zero current. \

It can be seen from Figure \ref{dwell} that the experimental data determine %
$I_{c1}$ quite accurately, within a few percent, because we don't
have to extrapolate very far from the experimental region to reach
the prefactor curve. \ On the other hand, this procedure clearly
will not work for determining  $I_{c2}$, because we are
extrapolating from positive to negative current, and a tiny change
in assumed slope causes a huge change in $I_{c2}$. \ Thus we
determine  $I_{c2}$ from Eq. \ref{Icratio} instead, and then
adjust $S_{2}$ to give the right value of $\tau _{2}$ at the
crossing point. \ The inset to Fig. \ref{dwell} shows that this
gives good semiquantitative agreement with the experimental data.
\ Although we forced the slope of $\tau _{1}$ to agree, the fact
that the slope of $\tau _{2}$ is much smaller is a true prediction
of the theory.

In addition to the room temperature data we fit in Fig. \ref{dwell}%
, Urazhdin \textit{et al}\cite{urazhdin} also measured dwell times at $T=4$%
K. \ We show this data in Fig. \ref{dwell} but it cannot be fit
well by the
theory. \ The reason for this can be seen graphically -- because the slopes of  $%
\tau _{1}$ and  $\tau _{2}$\ are similar and fairly large, both
will intersect the prefactor line at positive current, which is
inconsistent with the model.\  It has been
suggested\cite{urazhdin} that an effective LL noise temperature
which is different in the two wells (this could be due to Joule
heating or spin-wave excitation) could explain this, but the
present graphical construction suggests that this is not possible
-- some other mechanism must be involved.

The theory developed here is also applicable to the calculation of
magnetic noise in read heads\cite{microwave};
simulations\cite{periodic} of such systems show large, apparently
chaotic fluctuations under some circumstances, which are predicted
by the present theory as $I \rightarrow I_{c1}$.

\begin{acknowledgments}
This work was partially supported by NSF grants ECS-0085340 and
DMR-MRSEC-0213985, and by the DOE Computational Materials Sciences
Network.
\end{acknowledgments}

\newpage

\appendix

\section{EPAPS Supplementary Material}

\subsection{Basics of the Fokker-Planck (FP) equation}
The Kramers approach to chemical rate theory was adapted to the
magnetic switching problem by Brown\cite{brown}, who wrote a FP
equation for a probability density $\rho (M,t)$ on the sphere of
possible values for the magnetization $M$ (its magnitude is
assumed constant at its saturation value $M_{s}$). \ In magnetic
systems, the role of ''friction'' is played by the Landau-Lifshitz
damping coefficient $\alpha $. \ Physically occurring values of
$\alpha $ are low enough (about 0.01 to 0.1) that the system
nearly follows a constant-energy contour (one of the closed orbits
of the undamped
system) and there is a slow diffusion in energy. \ Brown and others\cite%
{coffey} who have used the FP equation have dealt mostly with the
case of non-equilibrium switching, in which one must specify an
initial ensemble with all systems in one of the two potential
wells, and details of the construction of this initial ensemble
can strongly affect the resulting rate. \ If the damping is weak,
the rate may be slower than TST due to delay in reaching
equilibrium near the barrier (this can lead to a rate
proportional to the damping coefficient in this limit\cite{kramers}\cite%
{brown}) \ It is worth noting that the telegraph noise problem is
in some ways simpler, since we can deal with a steady-state
distribution, and the damping-independent TST rate is always a
good approximation for physical values of $\alpha $. \

If $\alpha $ is
large, we should use the Gilbert formulation\cite{general} and replace $%
\gamma $ by $\gamma /(1+\alpha^2)$, in Eqs. \ref{Mcons} and
\ref{MLL}, but here we take $\alpha $ to be small.

\subsection{Defining temperatures in a magnetic system}

In a magnetic system one must make distinctions among several
different temperatures. \ Clearly if one puts a high current
through a nanoscale magnetic element, there is the possibility of
Joule heating, making the lattice temperature of the element
higher than that of the substrate (considered as a heat sink). \
In the Fokker-Planck equation, another temperature is the
Landau-Lifshitz noise temperature, related to the diffusion
constant $D$ in the FP equation by the fluctuation-dissipation
theorem (Eq. \ref{fluctdis}). \ Our steady-state solution of the
FP equation allows us to relate this LL noise temperature $T$ to
the Maxwell-Boltzmann temperature, defined by
$1/k_{B}T_{MB}=\partial \ln \rho ^{\prime }(E)/\partial E$, where
$\rho ^{\prime }(E)$ is the probability distribution in energy, by
Eq. \ref{drho}, $T_{MB}=T/\left[ 1-\eta (E)J/\alpha M_{s}\right]
.$

\subsection{Effective field (Eq. \ref{gradd})}

\begin{figure}[tbh]
\begin{center}
\includegraphics[height=1.5 in]{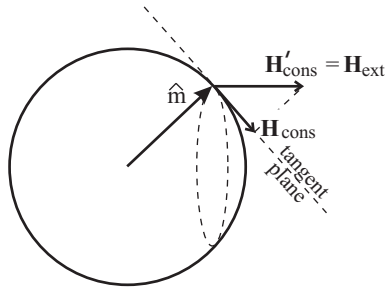}
\end{center}
\par
\vspace{-0.2 in} 
\caption{Magnetization vector for the simplest case of precession
about a horizontal external field, showing the
$\mathbf{M}$-sphere, the tangent
plane (perpendicular to the paper) and the projected field $\mathbf{H}_{%
\text{cons}}$. \ The dotted circle is the Stoner-Wohlfarth orbit,
a curve of constant energy.} \label{tan}
\end{figure}

The effective field is usually defined by
\begin{equation}
\mathbf{H}_{\text{cons}}^{\prime }=\mathbf{H}_{\text{ext}}+\mathbf{H}_{K}%
\hat{e}(\hat{\mathbf{m}}\cdot \hat{\mathbf{e}})-M_{s}\hat{\mathbf{z}}(\hat{%
\mathbf{m}}\cdot \hat{\mathbf{z}})  \label{A.Hcons}
\end{equation}%
but can in fact be defined differently (with the anisotropy field
perpendicular to the easy axis instead of along it, for example)
as long as the component perpendicular to $\mathbf{{\mathbf{M}}}$,
which we have denoted by $\mathbf{H}_{\text{cons}}$ (Fig.
\ref{tan}) and which can be
formally written as $\mathbf{H}_{\text{cons}}=-\hat{\mathbf{m}}\times (\hat{%
\mathbf{m}}\times \mathbf{H}_{\text{cons}}^{\prime })$, is
unchanged. \ The gradient $\mathbf{H}_{\text{cons}}$ (Eq.
\ref{gradd}) defined in the text is just the component
perpendicular to $\mathbf{{\mathbf{M}}}$ of the usual
formula (Eq. \ref{A.Hcons}) for the field, and the component along $\mathbf{{%
\mathbf{M}}}$ has no effect on the dynamics.

\bigskip

The directions of the various torques in the LL equation are shown
in insets to Fig. \ref{cont} of the text,which also shows the
contours of constant
energy on a planar projection of the $\mathbf{M}$-sphere. For in-plane $%
\mathbf{M}$ ($\theta =\pi /2)$, the directions are particularly
simple, as shown in the lower inset of Fig. \ref{cont} \ The
conservative (precession) term $\dot{\mathbf{M}}_{\text{cons}}$ is
vertical (along the energy contour,
\textit{i.e.}, the Stoner-Wohlfarth orbit), the LL damping term $\dot{%
\mathbf{M}}_{\text{LL}}$\ is horizontal (along the negative energy
gradient, \textit{i.e.}, $\mathbf{H}_{\text{cons}}$), and the
Slonczewski torque term is horizontal and opposite to the damping
term (for our choice of sign for the current).

For a general direction of $\mathbf{M}$, (upper inset) the
precession and LL damping torques are still exactly along and
perpendicular to the orbit, respectively. \ The Slonczewski torque
$\dot{\mathbf{M}}_{\text{Slon}}$, on the other hand, can be in any
direction, depending on the thick-layer magnetization direction.

\subsection{Derivation of Fokker-Planck equation}

\ Inserting Eq. \ref{j} for the current $\mathbf{j}(\mathbf{M},t)$
into the continuity equation (\ref{contin}) gives the FP equation

\begin{equation}
\frac{\partial \rho }{\partial t}=D\nabla ^{2}\rho -\nabla \cdot \mathbf{%
(\rho }\dot{\mathbf{M}}_{\text{det}})\text{.}  \label{A.FP}
\end{equation}%
This is identical to the FP equation derived by Brown\cite{brown},
though the latter looks more complicated because the 2D Laplacian
is written explicitly in spherical coordinates. \ It is important
that the divergence is two-dimensional here. If a
three-dimensional divergence is used, it has been shown [J. L.
Garcia-Palacios and F. J. Lazaro, "Langevin dynamics study of the
dynamical properties of small magnetic particles", Phys. Rev. B
\textbf{58}, 14937 (1998)] that the It\^{o} interpretation of the
LL equation gives an extra (radial) term in the probability
current J, relative to the Stratonovich interpretation. By working
entirely in the spherical surface, we ensure that the It\^{o} and
Stratonovich results are the same.

\subsection{Derivation of energy current}

\begin{figure}[tbh]
\begin{center}
\includegraphics[height=1.25 in]{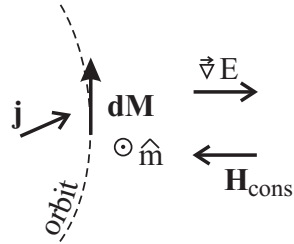}
\end{center}
\par
\vspace{-0.2 in} 
\caption{A length element $d\mathbf{M}$ along an orbit, for
derivation of the energy-current equation ((\ref{jE})). \ The
plane of the figure is the tangent plane to the
$\mathbf{M}$-sphere.} \label{dM}
\end{figure}

By definition of the 2D current $\mathbf{j}(\mathbf{M},t)$, the
rate at which systems cross the length element $d\mathbf{M}$ (Fig.
\ref{dM}) along
the contour (\textit{i.e.}, Stoner-Wohlfarth orbit) is the component of $%
\mathbf{j}$ perpendicular to the element.
Thus the total rate of crossing from lower to higher $E$ is an
integral over the orbit:
\begin{equation}
j^{E}_{i}(E,t)=\oint \left[ \mathbf{j}(\mathbf{M},t)\times
d\mathbf{M}\right] \cdot \hat{\mathbf{m}}  \label{A.jEdef}
\end{equation}

Using Eqs. \ref{j} and \ref{Mdet} for $\mathbf{j}$, we obtain

\begin{equation}
j^{E}_{i}(E,t)=\oint \left[ \rho (\mathbf{M},t)\dot{\mathbf{M}}_{\text{det}}(%
\mathbf{M})\right] \cdot \hat{\mathbf{m}}-\oint \left[ D\nabla \rho (\mathbf{%
M},t)\times d\mathbf{M}\right] \cdot \hat{\mathbf{m}}
\label{A.jE}
\end{equation}

The current from the conservative term (Eq. \ref{Mcons}) is along $d\mathbf{M%
}$ and does not contribute to $j^{E}_{i}$.\ The current (omitting
the subscript $i$ since we deal with only one well) then has three
terms:
\begin{equation}
j^{E}(E,t)=j^{E}_{\text{LL}}(E,t)+j^{E}_{\text{Slon}}(E,t)+j^{E}_{\text{diff}%
}(E,t)  \label{jEsum}
\end{equation}

The first (Landau-Lifshitz damping) term comes from the
Landau-Lifshitz
damping torque $\dot{\mathbf{M}}_{\text{LL}}$ (Eq. \ref{MLL}):%
\begin{equation}
j_{\text{LL}}^{E}(E,t)=\gamma \alpha M_{s}\oint \left[ \rho (\mathbf{M},t)%
\mathbf{H}_{\text{cons}}\times d\mathbf{M}\right] \cdot \hat{\mathbf{m}}%
=-\gamma \alpha M_{s}\rho ^{\prime }(E,t)I_{i}^{E}(E)
\label{A.jLL}
\end{equation}%
where we have used the fact that the energy is constant over the
orbit to bring $\rho $ (Eq. \ref{rhoE}) out of the integral, and
defined a (positive) energy integral $I_{i}^{E}$ by
\begin{equation}
I_{i}^{E}(E)\equiv \oint \left[ d\mathbf{M}\times \mathbf{H}_{\text{cons}}%
\right] \cdot \hat{\mathbf{m}}=\oint H_{\text{cons}}dM  \label{Ii}
\end{equation}%
We obtain the last expression because the three vectors in the
triple product are mutually orthogonal.

The remaining term in the deterministic torque is the Slonczewski torque%
\begin{equation}
\dot{\mathbf{M}}_{\text{Slon}}=-\gamma JM_{s}\hat{\mathbf{m}}\times (\hat{%
\mathbf{m}}\times \hat{\mathbf{m}}_{\text{p}})=-\gamma JM_{s}\left[ \hat{%
\mathbf{m}}(\hat{\mathbf{m}}\cdot \hat{\mathbf{m}}_{\text{p}})-\hat{\mathbf{m%
}}_{\text{p}}\right]
\end{equation}%
which gives an energy current%
\begin{equation}
j_{\text{Slon}}^{E}(E,t)=-\gamma JM_{s}\oint \left[ \rho
(\mathbf{M},t)\left[
\hat{\mathbf{m}}(\hat{\mathbf{m}}\cdot \hat{\mathbf{m}}_{\text{p}})-\hat{%
\mathbf{m}}_{\text{p}}\right] \times d\mathbf{M}\right] \cdot \hat{\mathbf{m}%
}=\gamma JM_{s}\hat{\rho ^{\prime }(E,t)\mathbf{m}}_{\text{p}}\cdot \oint %
\left[ d\mathbf{M}\times \hat{\mathbf{m}}\right]   \label{A.jSlon}
\end{equation}%
since the first term in the cross product is orthogonal to $\hat{\mathbf{m}}$.%
\ Defining the magnetization integral
\begin{equation}
\mathbf{I}_{i}^{M}(E)=\oint d\mathbf{M}\times \mathbf{M}
\label{MI}
\end{equation}%
gives
\begin{equation}
j_{\text{Slon}}^{E}(E,t)=\gamma J\hat{\rho ^{\prime }(E,t)\mathbf{m}}_{\text{%
p}}\cdot \mathbf{I}_{i}^{M}  \label{jSlon}
\end{equation}

The last (diffusive) term in Eq. \ref{A.jE} involves
\begin{equation}
\nabla \rho (\mathbf{M},t)=\nabla \rho ^{\prime }(E(\mathbf{M}),t)=\frac{%
\partial \rho ^{\prime }(E,t)}{\partial E}\nabla E(\mathbf{M})=-\frac{%
\partial \rho ^{\prime }(E,t)}{\partial E}\mathbf{H}_{\text{cons}}
\end{equation}%
\label{A.gradrho} and gives the same triple vector product as the
Landau-Lifshitz damping term (Eq. \ref{A.jLL}):
\begin{equation}
j^{E}_{\text{diff}}(E,t)=D\frac{\partial \rho ^{\prime }(E,t)}{\partial E}%
\oint \left[ \mathbf{H}_{\text{cons}}\times d\mathbf{M}\right] \cdot \hat{%
\mathbf{m}}=-D\frac{\partial \rho ^{\prime }(E,t)}{\partial
E}I_{i}^{E}(E); \label{A.jdiff}
\end{equation}

Our final result for the total energy current (Eq. \ref{jEsum}) is thus Eq. %
\ref{jE}. \ Inserting this current into the continuity equation
(Eq. \ref{Econtin}) gives the energy Fokker-Planck equation. \

Since Kramers used the FP equation in energy only in the
low-damping limit, which might suggest that it is an approximation
valid only in that limit, it is important to realize that it is
not an approximation at all -- as long as the probability density
$\rho$ is constant along an orbit, it gives the exact evolution of
$\rho(E,t)$.

In a steady state, the energy current (Eq. \ref{jE}) vanishes,
giving Eq. \ref{drho}:
\begin{equation}
\frac{\partial \ln \rho ^{\prime }(E)}{\partial E}=\frac{\gamma
}{D}\left[ -\alpha M_{s}+\eta (E)J\right] \equiv -V\beta (E)
\label{A.drho}
\end{equation}%
This involves the ratio (Eq. \ref{etadef})
\begin{equation}
\eta (E)=\frac{\hat{\mathbf{m}}_{\text{p}}\cdot \mathbf{I}_{i}^{M}(E)}{%
I_{i}^{E}(E)}. \label{A.etadef}
\end{equation}%
plotted in Fig. \ref{eta}.  We have calculated the integrals by
solving the LL equation numerically at each energy; an analytic
check is possible at the bottom of each well (at the left end of
the $\eta_1$ or $\eta_2$ curve), where $| \eta_i(E_i) | =%
M_s / [\frac{1}{2}M_{s}+H_{K}\mp H_{\text{ext}}]\sim 2$.

\subsection{Solving the FP equation in energy}

We can write the inverse effective Maxwell-Boltzmann temperature
defined by Eq. \ref{A.drho} or \ref{drho} as
\begin{equation}
\beta_i (E)=\frac{1}{k_B T_{MB}}=%
\frac{1}{k_{B}T}[1-\eta_i (E)J/\alpha M_{s}]
\end{equation}%
in terms of which Eq. \ref{rho} can be written
\begin{equation}
\rho _{i}^{\prime }(E)=\rho ^{\prime }(E_{\text{sad}})\exp \left[
V\int\limits_{E}^{E_{\text{sad}}}\beta _{i}(E^{\prime })dE^{\prime
}\right] \label{A.rho}
\end{equation}
Note that $\rho ^{\prime }(E_{\text{sad}})$ has no region index
because it is the same in all three regions.  If $\beta_i(E)$ is
independent of energy, Eq. \ref{A.rho} becomes the usual
Maxwell-Boltzmann distribution; in any event it is easy to
integrate numerically.

Current-driven thermal switching can be understood in terms of the
increase in the Arrhenius rate due to this temperature increase. \
If we increase the current $J$ enough, the temperature at the
bottom of the well can be made negative; this can be used to model
the onset of microwave noise, although our first-order treatment
of $J$ will eventually fail and we will need to work in terms of
nonzero-$J$ orbits rather than the present unperturbed orbits.

\subsection{Derivation of TST (transition state theory) rate}

\begin{figure}[tbh]
\begin{center}
\includegraphics[height=1.2 in]{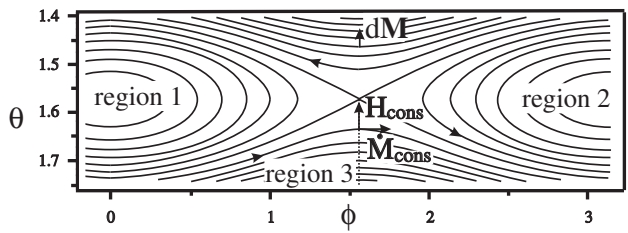}
\end{center}
\par
\vspace{-0.2 in} 
\caption{The contour plot of Fig. \ref{cont}, showing the dashed
line across which switching from region 2 to region 1 takes place,
along which we integrate the current to obtain the total TST rate.
} \label{TST}
\end{figure}

The TST rate is the probability per unit time of crossing the
dotted line in Fig. \ref{TST}, which follows the energy gradient
from the saddle point.
This is
independent of the damping coefficient, since the damping current %
$\rho \dot{\mathbf{M}}_{\text{LL}}$ (Eq. \ref{MLL}) %
has no component normal to this line, nor does the diffusive current %
$-D\nabla \rho $, and the conservative and Slonczewski currents
are independent of damping. \ To lowest order in the electric
current $J$, we need only include the conservative current, and
the probability current (left to right) is given by an integral
like Eq. \ref{A.jE}:
\begin{equation}
j_{\text{TST}}=-\oint \left[ \mathbf{j}_{\text{cons}}(\mathbf{M})\times d%
\mathbf{M}\right] \cdot \hat{\mathbf{m}}=-\oint \left[ \dot{\rho \mathbf{M}}%
_{\text{cons}}\times d\mathbf{M}\right] \cdot \hat{\mathbf{m}}
\label{A.TST1}
\end{equation}%
[Since we are in a steady state, the reverse current (right to
left, across a similar line pointing upward from the saddle point
in Fig. \ref{TST}) is equal in magnitude.]  Using Eq. \ref{Mcons},
and observing from Fig. \ref{TST} that all the cross products
involve perpendicular vectors, we obtain

\begin{equation}
j_{\text{TST}}=\gamma M_{s}\oint \rho _{3}^{\prime }(E)H_{\text{cons}}dM=%
\frac{\gamma M_{s}}{\mu _{0}}\int_{E_{\text{sad}}}^{\infty }\rho
_{3}^{\prime }(E)dE  \label{A.TST2}
\end{equation}%
where we have used $dE=\mu _{0}H_{\text{cons}}dM$ (Eq.
\ref{gradd}). \ The analog of Eq. \ref{rho} for the region
$E>E_{\text{sad}}$ where we integrate
up instead of down from $E_{\text{sad}}$) is%
\begin{equation}
\rho _{3}^{\prime }(E)=\rho ^{\prime }(E_{\text{sad}})\exp \left[ -\frac{V}{%
k_{B}T}[1-\overline{\eta _{3}}(E)J/\alpha
M_{s}][E-E_{\text{sad}}]\right] \label{A.rho3}
\end{equation}%
so that%
\begin{equation}
j_{\text{TST}}=\frac{\gamma M_{s}}{\mu _{0}}\rho ^{\prime }(E_{\text{sad}%
})\int_{E_{\text{sad}}}^{\infty }\exp \left[ -\frac{V}{k_{B}T}[1-\overline{%
\eta _{3}}(E)J/\alpha M_{s}][E-E_{\text{sad}}]\right] dE
\label{A.TST3}
\end{equation}%
Noting from Fig. \ref{eta} that $\overline{\eta _{3}}$ is slowly
varying within $k_{B}T$ of $E_{\text{sad}}$, we approximate it by
a constant so the integral is analytic:

\begin{equation}
j_{\text{TST}}=\frac{\gamma M_{s}k_{B}T}{\mu _{0}V[1-\overline{\eta _{3}}%
(E)J/\alpha M_{s}]}\rho ^{\prime }(E_{\text{sad}})  \label{A.jTST}
\end{equation}%
It may seem inconsistent to include a term in $J/\alpha $ when we
are working to lowest order in $J$; however, since $\alpha $ is
also small, we really work to lowest order in both, and $J/\alpha
$\ is of zeroeth order in this sense.

In the chemical reaction rate literature\cite{chandler} it is
found that TST gives a good representation of the switching rates
as long as a trajectory that crosses the saddle point from the
initial to the final well is likely to be trapped there by losing
energy to LL damping. \ The LL damping constants in magnetic
materials are large enough (0.01-0.02) that escape after one orbit
in the final well is quite unlikely, so the TST is likely to be
quite accurate. \ Our telegraph-noise fitting involves rates that
vary by many orders of magnitude, so corrections to TST of a few
percent are not critically important.

\subsection{Derivation of dwell times}

The total probability of being in well $i$ (= 1 or 2) is obtained
by integrating the density over the well. \ The probability of
being in a surface element $d^{2}M$ on the M-sphere with energy
$E$ is $\rho ^{\prime
}(E)d^{2}M$, so the probability of being in the ring shown in Fig. \ref%
{A.ring} is
\begin{figure}[tbh]
\begin{center}
\includegraphics[height=1.25 in]{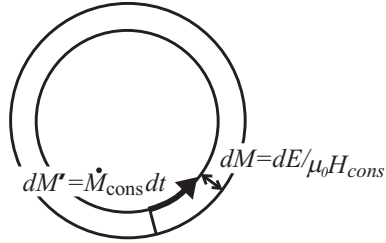}
\end{center}
\par
\vspace{-0.2 in} 
\caption{A ring on the M-sphere representing energies between $E$
and $E+dE$, showing the dimensions $dM$ and $dM^{\prime }$ of an
area element.} \label{A.ring}
\end{figure}
\begin{equation}
\int \rho ^{\prime }(E)\dot{\mathbf{M}}_{\text{cons}}dtdE/\mu _{0}H_{\text{%
cons}}=\gamma \rho ^{\prime }(E)\int M_{s}H_{\text{cons}}dtdE/\mu _{0}H_{%
\text{cons}}=\frac{\gamma M_{s}P_{i}(E)}{\mu _{0}}\rho ^{\prime
}(E)dE \label{A.DOS}
\end{equation}%
where $P_{i}(E)$ is the period of the orbit. Thus the well
probability is (using Eq. \ref{A.rho3} for $\rho _{3}^{\prime }$)
\begin{equation}
p_{i}=\int_{E_{i}}^{E_{\text{sad}}}\frac{\gamma M_{s}P_{i}(E)}{\mu
_{0}}\rho _{i}^{\prime }(E)dE=\rho ^{\prime
}(E_{\text{sad}})\frac{\gamma M_{s}}{\mu
_{0}}\int_{E_{i}}^{E_{\text{sad}}}P_{i}(E)\exp \left[ \frac{V}{k_{B}T}[1-%
\overline{\eta _{i}}(E)J/\alpha M_{s}][E_{\text{sad}}-E]\right] dE
\label{A.p0}
\end{equation}%
where $E_{i}$ is the energy at the bottom of well $i$. \ Again, $\overline{%
\eta _{i}}(E)$ and $P_{i}(E)$ are nearly independent of $E$, which
is true
in the lowest few $k_{B}T$ of the well, so this integrates to%
\begin{equation}
p_{i}=\frac{\gamma M_{s}P_{i}(E_{i})}{\mu _{0}V}\frac{\rho ^{\prime }(E_{%
\text{sad}})k_{B}T}{1-\overline{\eta _{i}}(E_{i})J/\alpha M_{s}}\left[ \exp %
\left[ \frac{V}{k_{B}T}[1-\overline{\eta _{i}}(E_{i})J/\alpha M_{s}]E_{i}^{b}%
\right] -1\right]   \label{A.p}
\end{equation}%
Then the dwell time is
\begin{equation}
\tau
_{i}=\frac{p_{i}}{j_{\text{TST}}}=P_{i}(E_{i})\frac{1-\overline{\eta
_{3}}(E_{\text{sad}})J/\alpha M_{s}}{1-\overline{\eta
_{i}}(E_{i})J/\alpha
M_{s}}\left[ \exp \left[ \frac{V}{k_{B}T}[1-\overline{\eta _{i}}%
(E_{i})J/\alpha M_{s}]E_{i}^{b}\right] -1\right]   \label{A.dwell}
\end{equation}%
where $E_{i}^{b}\equiv E_{\text{sad}}-E_{i}$ is the barrier
height. \ The important thing to notice is that $\overline{\eta
_{1}}$ and $\overline{\eta _{2}}$ have opposite signs, so one
dwell time increases exponentially with $J $ and the other
decreases, as in Fig. \ref{dwell}.

\subsection{Determination of energy barriers}
One virtue of the fitting scheme described in the text and
illustrated in Fig. \ref{dwell} is that we can determine the
stability factor $S_{i}\equiv VE^b_{i}/k_{B}T_{i}$ of each well
independently, where the barriers are given by\cite{brown}
\begin{equation}
E^b_{i}=(1\mp H_{\text{ext}}/H_{K})^{2}KV/k_{B}T_{i} \label{A.Eb}
\end{equation}%
though these are in turn consistent with many choices of
anisotropy $K=H_K M_s / 2$, effective external field
$H_{\text{ext}}$, temperature $T_{i}$ , and volume $V$. \ For
example, if we assume the wells are at the same temperature and
$H_{\text{ext}}=-120$ Oe, we obtain
\begin{equation}
 H_{K}=220 \text{ Oe .}  \label{A.Hk}
\end{equation}%
If we further assume the experimental estimate\cite{urazhdin}
$V=3.67\times 10^{-23}$ m$^{3},$ we obtain an effective
temperature $T=730$ K. (Of course, this could be reduced to room
temperature by assuming a smaller effective volume.) \


\end{document}